\documentclass[a4paper,11pt]{article}
\usepackage{pos}

\newcommand{\sNN}{$\sqrt s_{\rm NN}$}
\newcommand{\s}{$\sqrt s$}
\newcommand{\gev}{GeV/$\it c$}
\newcommand{\pp}{$p$+$p$}
\newcommand{\AuAu}{Au+Au}
\newcommand{\RuRu}{Ru+Ru}
\newcommand{\ZrZr}{Zr+Zr}
\newcommand{\pT}{$p_{\mathrm{T}}$}
\newcommand{\pTjet}{$p_{\mathrm{T,jet}}$}
\newcommand{\DeltaR}{$\Delta R$}
\newcommand{\rr}{$R$}
\newcommand{\Zg}{$z_{\mathrm{g}}$}
\newcommand{\Rg}{$R_{\mathrm{g}}$}
\newcommand{\DMbyM}{$\Delta M/M$}
\newcommand{\rhoR}{$\rho (r)$}
\newcommand{\pizero}{$\pi^{0}$}
\newcommand{\gammadir}{$\gamma_{\mathrm{dir}}$} 
\newcommand{\ETtrig}{$E_{\mathrm T}^{\rm trig}$}

\newcommand{\Rbrtwofive}{\ensuremath{\mathfrak{R}^{0.2/0.5}}}
\newcommand{\RAA}{$R_{\mathrm{AA}}$}
\newcommand{\vTwo}{$v_{2}$}
\newcommand{\Npart}{$\langle  N_{\rm part} \rangle$}

\newcommand{\Jpsi}{$J/\psi$}

\title{STAR Experimental Highlights at Hard Probes 2023}
\author*[a,b,c]{Nihar Ranjan Sahoo (for the STAR collaboration)}
\affiliation[a]{National Institute of Science Education and Research, HBNI, Jatni 752050, India\\}
\affiliation[b]{
Cyclotron Institute, Texas A\&M University, College Station, USA\\}
\affiliation[c] {\it Institute of Frontier and Interdisciplinary Science, 
Shandong University, Qingdao, Shandong, 266237, China}

\emailAdd{nihar@rcf.rhic.bnl.gov}
\abstract{We highlight the STAR experiment's hard probes results, including jets and heavy flavor production in heavy-ion collisions to study the properties of the quark-gluon plasma. Various jet-substructure observables in proton-proton collisions are presented to explore both perturbative and non-perturbative regimes of quantum chromodynamics. Finally, we discuss the STAR experiment's hard probes physics program for ongoing data taking.
 } 

\FullConference{% HardProbes2023\\
 26-31 March 2023 \\
 Aschaffenburg, Germany\\}

\begin{document}
\maketitle

%%%________ Introduction__________
\section{Introduction}
The STAR experiment at the RHIC facility primarily aims to investigate the hot-dense state of quantum chromodynamics (QCD) matter--- quark-gluon plasma (QGP)---in order to accomplish the scientific objectives of the RHIC mission. 
Amassing a vast amount of data from different collision species and energies enables us to explore the properties of QGP under different conditions and QCD in vacuum at RHIC. 

In the context of heavy-ion collisions, jets and heavy flavor hadrons serve as important hard probes for studying QGP. Their production and interaction with the medium in heavy-ion collisions provide crucial information of QGP's transport properties, medium temperature, and other characteristics. 

Jets, which are collimated sprays of hadrons, emerge from scattering between two incoming partons with significant momentum transfer, commonly known as "hard scattering". The jet production rate is calculable using perturbative QCD (pQCD) and is compared to \pp\ data obtained at both RHIC and the LHC. However, the yield of jets is suppressed in heavy-ion collisions due to jet-medium interaction--a phenomenon known as jet quenching~\cite{Wang:1994fx}. This phenomenon serves as a distinctive signature of QGP. 
 
On the other hand, heavy flavor quarks--such as charm ($c$) or beauty ($b$) quarks or antiquarks--are produced mainly from hard scattering in both heavy-ion and \pp\ collisions. These heavy-flavors are observed in experiment either in the form of quarkonium ($c\bar{c}$ and $b\bar{b}$) states or open heavy flavor hadrons. They interact with the medium while traversing through it and their yield suppression in heavy-ion collisions stands as a key signature of QGP~\cite{STAR:2022rpk,Dong:2019byy}. 

%The suppression of heavy-flavor hadrons yield in heavy-ion collisions stands as a key signature of the QGP.

In addition, the study of the QCD evolution in vacuum and hadronization processes enables exploring the two regiems: pQCD and non-perturbative (npQCD). The transition between pQCD and npQCD is mainly controlled by the $\Lambda_{\mathrm{QCD}}$ parameter in the running coupling constant. The jet-substructure observables in \pp\ collisions at RHIC energy are investigated exploring these regimes and serving a baseline for the heavy-ion collisions as well.

We highlight the STAR experiment's key hard probes results at the Hard Probes 2023 conference in Aschaffenburg, Germany. To study the QCD evolution and hadronization in vacuum, the energy-energy correlator within jet and observables related to jet mass and jet shape are presented. In-depth analysis on jet quenching phenomenon in heavy-ion collisions is performed by measuring inclusive charged hadrons and jets recoiling from hadrons or direct photons in heavy-ion collisions. The heavy-flavor hadron and quarkonia suppression and their flow in QGP medium are also presented. 
A comprehensive perspective of QGP properties and the implications of these findings are thoroughly discussed. Finally, the STAR experiment's recent detector upgrade and data taking plan for the hard probes physics program are presented.

%%%________ 
\section{STAR Detector}
\label{Sec:STARdetector}
\begin{figure*}[htb!]
    \centering
    \includegraphics[width=0.7\textwidth]{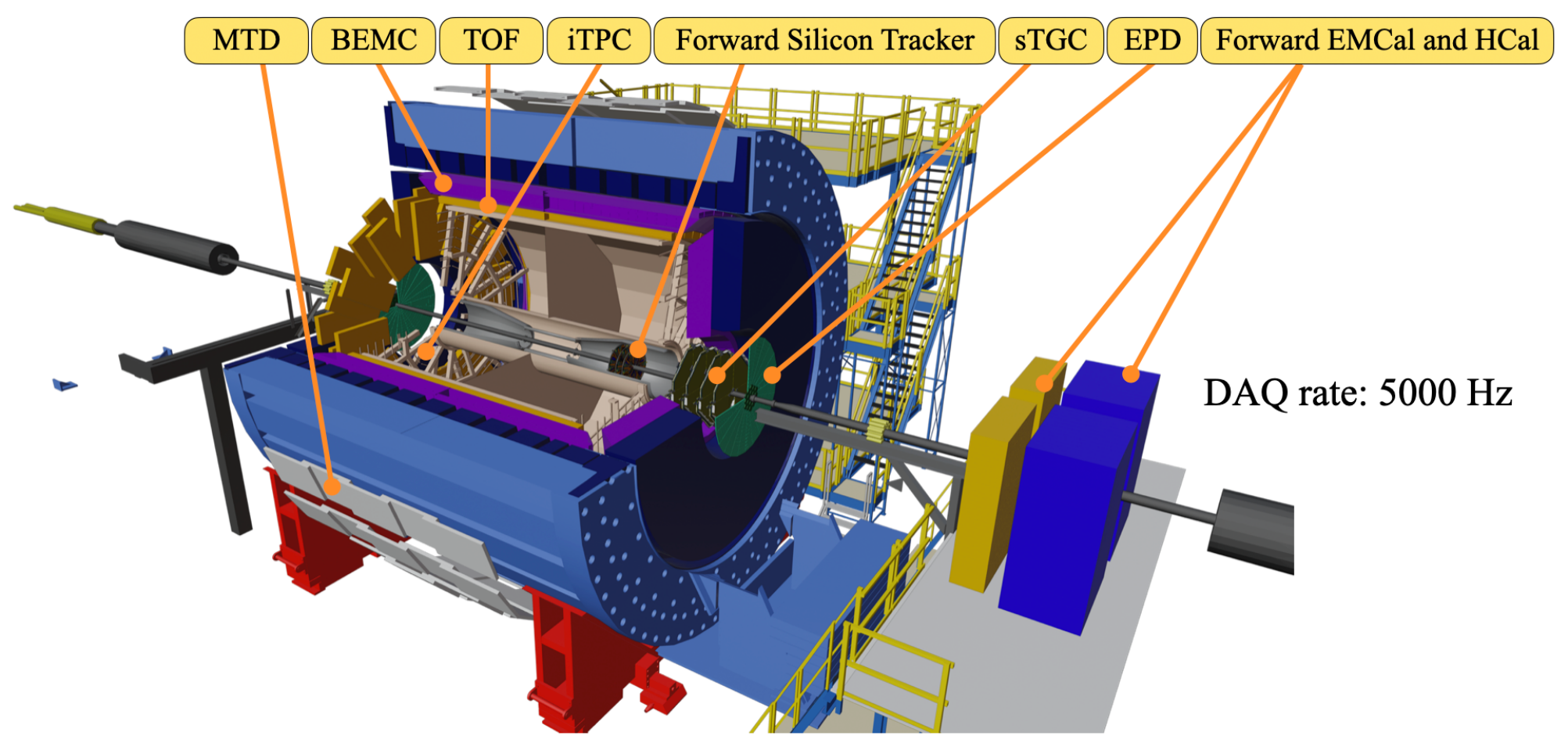}
    \caption{The STAR detector with the forward detector upgrade. }
    \label{Fig:STARdetector}
\end{figure*}

The STAR experiment comprises several detectors to measure particles in a 2$\pi$ azimuthal angle as shown in Fig.~\ref{Fig:STARdetector}. The Time Projection Chamber (TPC) is used to measure the charged hadrons and also for the particle identification (PID). A recent inner Time Projection Chamber (iTPC) detector upgrade helps to measure charged particles in psedorapidity range $|\eta|<$1.5 and transverse momentum  \pT\ $>$ 0.15 \gev. 
The time-of-flight (TOF) detector is used for PID in heavy-flavor analyses. 
The Barrel Electromagnetic Calorimeter (BEMC) is used for triggering, and for neutral particle measurement within 
$|\eta|<$1.  The Muon Telescope Detector (MTD) is situated outside the STAR magnet and serves the dual purpose of detecting high \pT\ muons and acting as the dimuon trigger. The two Event-Plane Detectors (EPD) are placed at both the east and west sides of STAR magnet covering 2.14 $< |\eta| <$ 5. The EPD is used for the event-plane determination pertaining to the flow measurements and also for the centrality determination in heavy-ion collisions.  

%The Event-Plane Detectors (EPD) are placed at both the east and west sides of STAR magnet covering 2.14 $< |\eta| <$ 5. 

In 2022, the STAR experiment underwent the forward detector upgrade. The four new subsystemts are the forward calorimeter system (FCS) consisting of the electromagnetic (EMCal) and hadronic calorimeter (HCal), and the forward tracking system (FST) with a silicon detector and a small strip Thin Gap Chambers tracking detector (sTGC) enabling detection of the neutral pions, photons, electrons, jets, and the charged hadrons within pseudorapidity range 2.5 $< \eta < $ 4. 

%The results presented in these highlights only include the data taken mainly using the BEMC, MTD, and TPC detectors. 

%_______________________ STAR detectot Fig.1 ______

%%%________ 
\section{Jet measurements in STAR} 
\label{Sec:Jetmeasurement}
%We present several jet measurements both in \pp\ and heavy-ion collisions. 
In this section, the jet substructure observables are discussed to study the QCD evolution and hadronization at RHIC energy. Different manifestations of jet-medium interactions in heavy-ion collisions are discussed below. 
%%%%%%%%%%%%%%%%%%%%%_____________
\subsection{Probing fundamental QCD: theory vs reality}
\label{Sec:Jetsubstructure}

{\it Energy-Energy Correlator}: The two-point energy correlator (or EEC) helps to study different regimes of QCD evolution from quark/gluon to hadrons as a function of their angular scale. In conformal field theory approach, the EEC allows one to measure the flow of energy in QCD~\cite{Chen:2020vvp,Komiske:2022enw}. The EEC is defined as a weighted distribution of the products of the jet energy fractions carried by all possible two constituent combinations within a jet and its expression can be found in Ref.~\cite{AndrewHP2023}.
%_______________________ EEC Fig.2 ______
\begin{figure*}[htb!]
    \centering
    \includegraphics[width=0.45\textwidth]{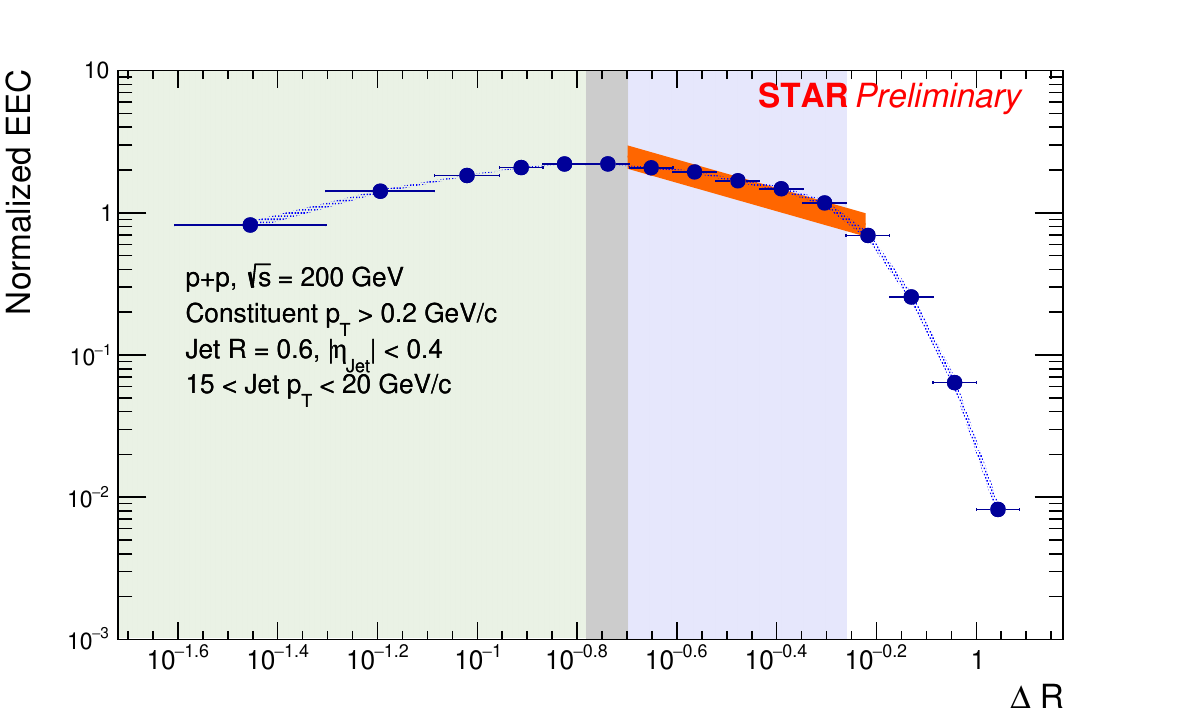}
    \caption{Corrected distributions of the normalized EEC plotted in \DeltaR\ for R = 0.4 for 15 < \pTjet < 20 \gev\ in \pp\ collisions at \s\ = 200 GeV. The free-hadron regime, transition region, and quark-and-gluon regime are highlighted in green, gray and purple respectively. The NLL-pQCD calculations are presented for (3 GeV)/\pTjet < \DeltaR < R. }
    \label{Fig:EEC}
\end{figure*}

The jets are reconstructed using anti-$\rm k_{T}$ algorithm in \pp\ collisions in \s\ = 200 GeV data. The jets are measured with jet resolution parameters \rr\ = 0.4 and 0.6; transverse momentum of jet within 15 < \pTjet\ < 20 \gev\ and 30 < \pTjet\ < 50 \gev. Figure~\ref{Fig:EEC} shows the normalized EEC observable as a function of angular scale \DeltaR\ for \rr\ = 0.4, and 15 < \pTjet\ < 20 \gev. Here \DeltaR\ is angular separation between the two constituents within a jet. The free hadron phase at small opening angles, perturbative behavior of quark/gluon phase at large opening angles, and transition region in between are shown as three colored regions. The Next-to-Leading-Log (NLL) pQCD calculation is comparable to the data at the quark-gluon regime. The value of \DeltaR\ \pTjet\ is between 2-3 GeV and this is independent of \pTjet\ range. This suggests that the confinement of quark/gluon degrees of freedom into hadrons occurs at universal momentum scale. A detailed discussion on this observable is in Ref.~\cite{AndrewHP2023}. \\

%_____________
{\it SoftDrop and CollinearDrop in \pp\ }: The SoftDrop technique~\cite{Larkoski:2014wba} is employed to effectively eliminate soft-wide angle radiations within a jet, thereby reducing contamination from initial state radiation, underlying event, and multiple hadron scattering in \pp\ collisions.

Two key observables resulting from the SoftDrop technique are the shared momentum fraction (\Zg) and the groomed radius (\Rg). These observables represent the momentum and angular scale, respectively, and can be utilized to map the Lund Plan of parton shower~\cite{Dreyer:2018nbf}.

To further examine the groomed soft components of a jet obtained through the SoftDrop technique, the CollinearDrop technique is employed. This technique utilizes the two specific kinematic settings from the SoftDrop approach, as discussed in Ref.~\cite{Chien:2019osu}. Using both these techniques, the soft and hard component of a jet can be examined to explore the pQCD and hadronization processes. 

\begin{figure*}[htb!]
    \centering
    \includegraphics[width=0.4\textwidth]{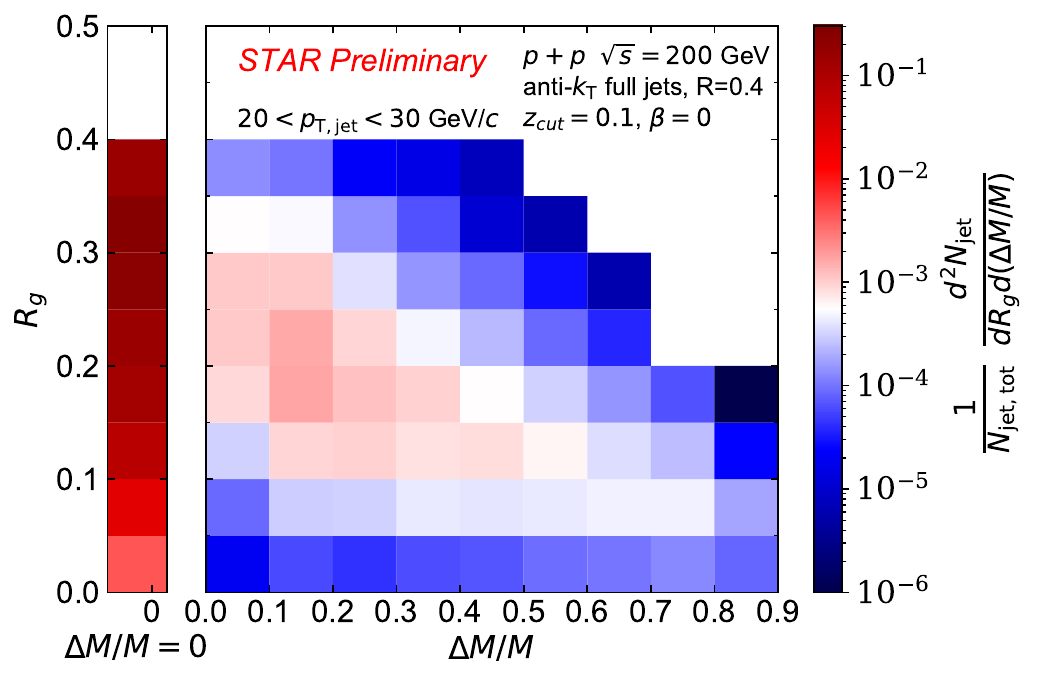}
     \includegraphics[width=0.3\textwidth]{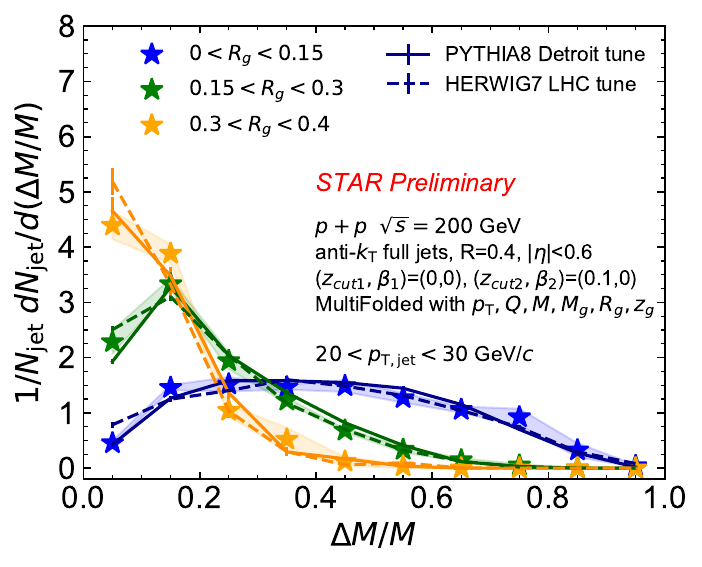}
    \caption{Correlation between \DMbyM\ and \Rg\ (left) unfolded with MultiFold method and the projection of \DMbyM\ for three different \Rg\ selections (right) for jets with \rr\ = 0.4 in \pp\ collisions at
\s\ = 200 GeV.}
    \label{Fig:JetSubstruct}
\end{figure*}

The \DMbyM\ observable is measured for different \Rg\ for jets with \rr\ = 0.4 in \pp\ collisions with 10 < \pTjet\ < 30 \gev\ as shown in Fig~\ref{Fig:JetSubstruct}. %Here \DMbyM\ represents the jet mass fraction resulting from the CollinearDrop as discussed in ~\cite{MonikaHP2023}. 
Here the \DMbyM\ represents the ratio between the difference of ungroomed and SoftDrop groomed over the ungroomed jet mass. It is corrected for the detector effects using MultiFold---a machine learning technique that preserves the correlations between multiple observables in unbinned method. It is observed that the \DMbyM\ is anti-correlated with \Rg\ corroborating the angular ordering of the parton shower. 
In addition, the groomed mass fraction ($\mu$), as defined and results shown in Ref.~\cite{MonikaHP2023}, is measured for different \Rg\ ranges in \pp\ collisions. The $\mu$ represents the mass sharing of the hard splitting; this decreases for narrower splits of a groomed jet.

%Decreasing trend of $\mu$ values o smaller values at smaller \Rg\ indicates reduction of virtuality of groomed jet with narrower splitting.

%A shift of $\mu$ to smaller values at smaller \Rg\ indicates a faster reduction of virtuality in the parton shower.\\

%Shift of μ to smaller values at smaller angles indicates a faster reduction of virtuality in the jet shower 

%Anti-correlation between collinear dropped jet mass ∆M/M and Rg → consistent with angular ordering of the parton shower

%_______________________ SoftDrop and CollinearDrop Fig.2 ______

\begin{figure*}[htb!]
    \centering
    \includegraphics[width=0.4\textwidth]{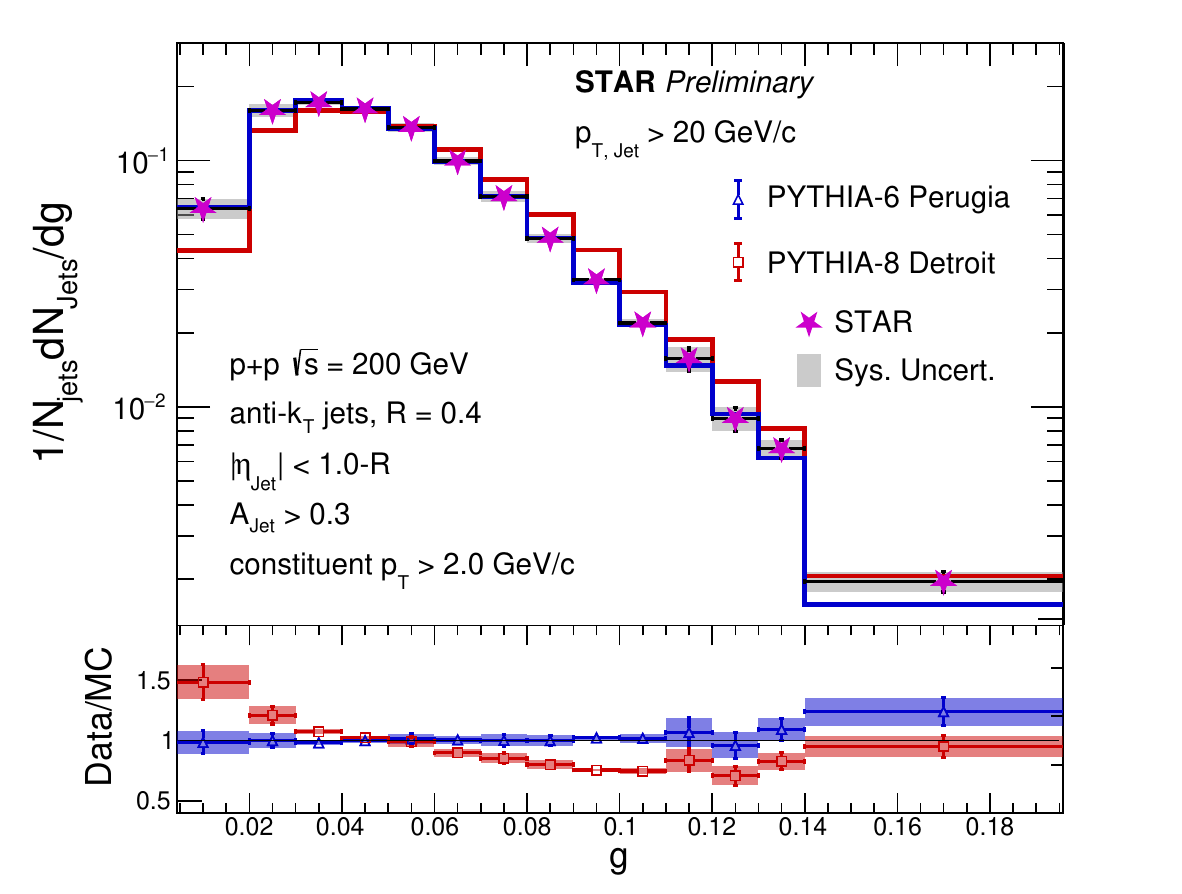}
     \includegraphics[width=0.3\textwidth]{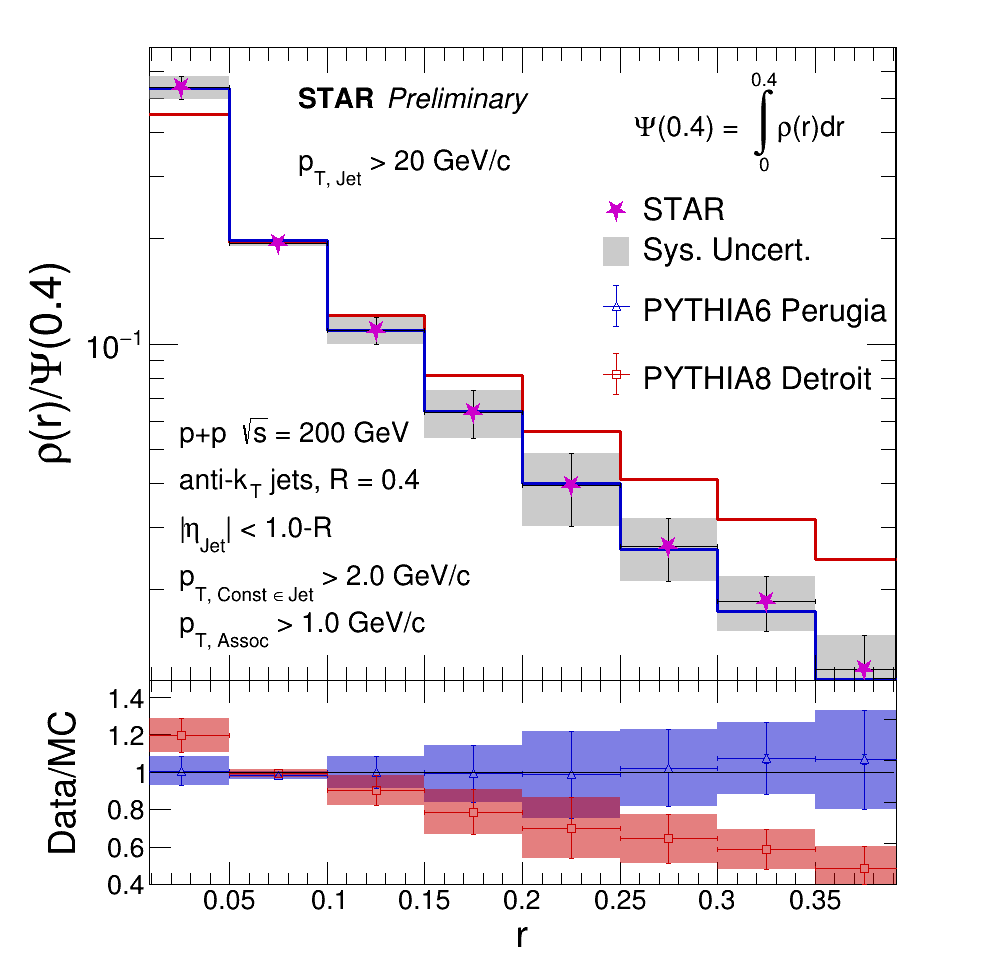}
    \caption{The $g$ (left) and \rhoR\ (right) distributions for jets with \rr\ = 0.4 and \pTjet\ > 20 \gev\ in \pp\ collisions at \s\ = 200 GeV.}
    \label{Fig:JetShap}
\end{figure*}

%%%%%%%%%%%%%%%%%%
{\it Generalized angularities and differential jet shapes in \pp}: We also report different jet shape observables, like jet-girth ($g$), momentum dispersion ($D$) and the differential jet-shape (\rhoR) as defined in ~\cite{TanmayHP2023}, in \pp\ collisions at \s\ = 200 GeV. Figure~\ref{Fig:JetShap} shows the $g$ and \rhoR\ distributions and their comparisons with different versions of PYTHIA. The data agree with the PYTHIA-6 Perugia tune~\cite{STAR:2019yqm} whereas PYTHIA-8 Detroit tune~\cite{Aguilar:2021sfa} needs further tuning for these observabels related to jet shape and momentum distributions inside jet. 
These jet shape obervables set baselines for the similar measurements in heavy-ion collisions to study the in-medium jet shape modification at RHIC. 

%Data agree with PYTHIA6 Perugia tune; PYTHIA-8 Detroit needs further tuning
%_______________________ Jet shape Fig.4 ______

%%%%%%%%%%%___________________
\subsection{Jet-medium interaction in heavy-ion collisions}
\label{Sec:Jetmediuminteraction}

{\it Intra-jet broadening and jet acoplanarity in central \AuAu\ collisions}: The interaction between a highly virtual parton and the medium involves the simultaneous effects of both vacuum shower and in-medium gluon radiation. This interplay plays a pivotal role in understanding the phenomena of jet-medium interaction. As a consequence of jet energy loss, the out-of-cone radiations contribute to intra-jet broadening in heavy-ion collisions. 

\begin{figure*}[htb!]
    \centering
    \includegraphics[width=0.4\textwidth]{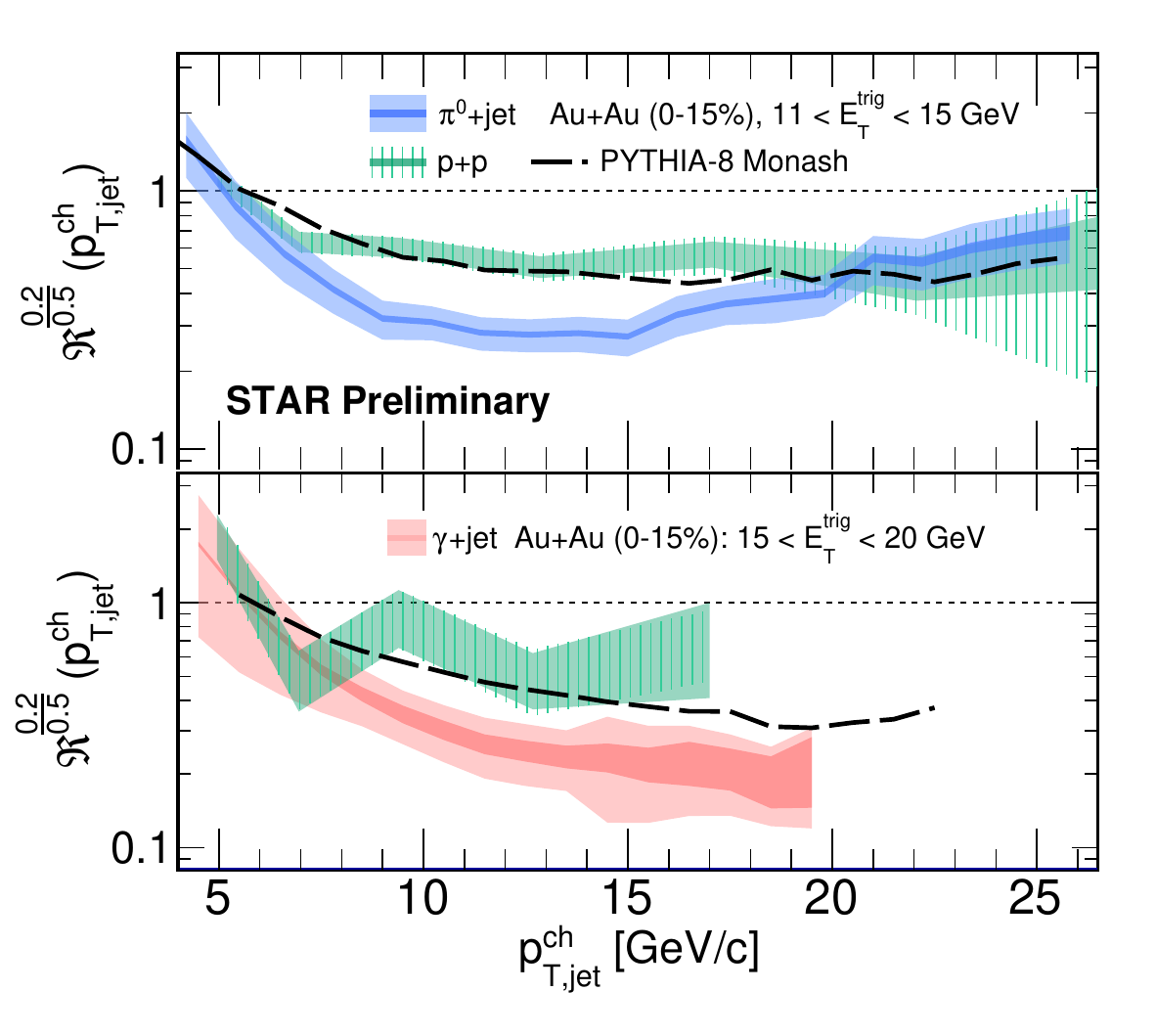}
     \includegraphics[width=0.38\textwidth]{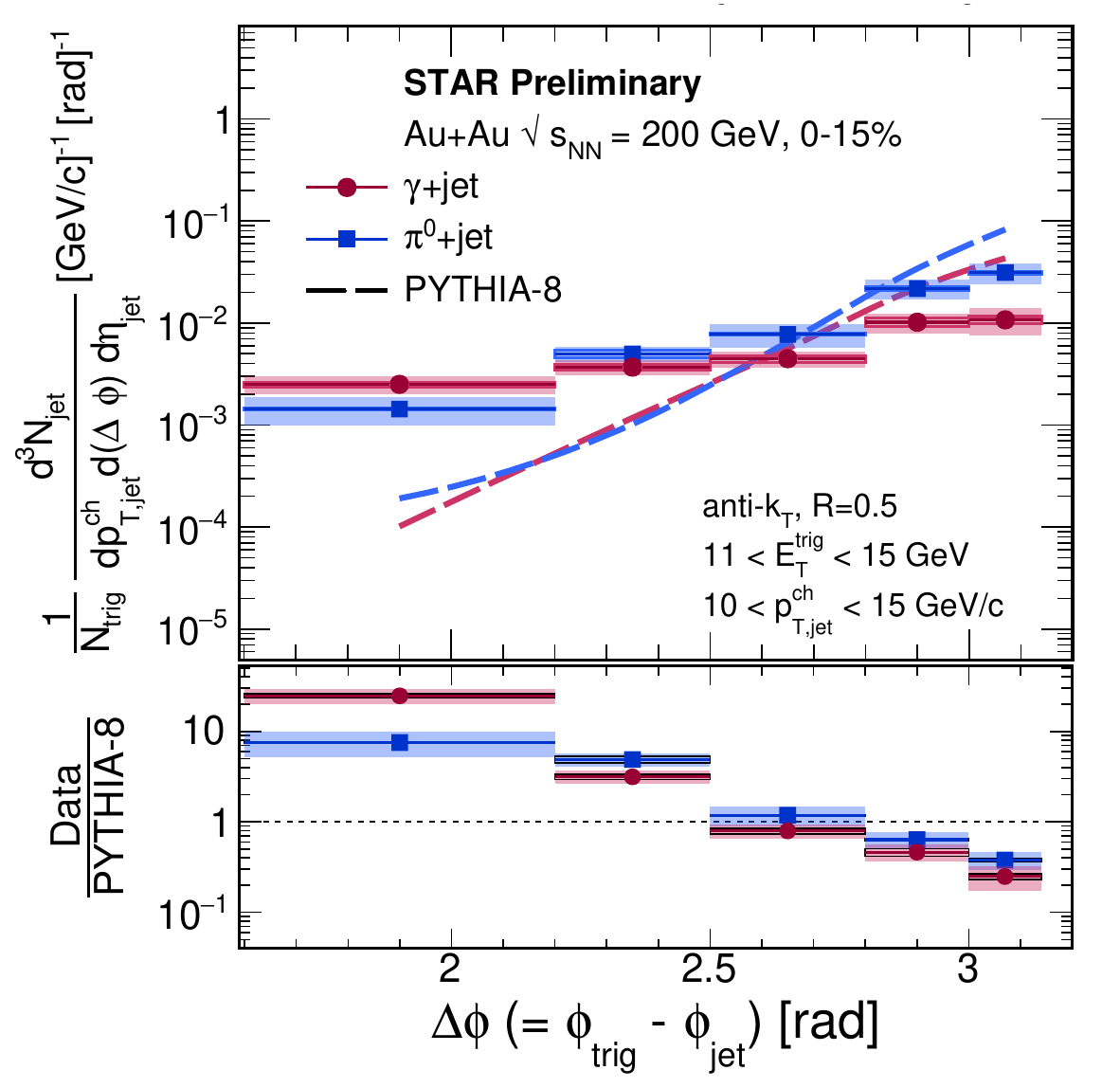}
    \caption{Left: Recoil jet yield ratios between \rr\ = 0.2 and 0.5 for \pizero +jet and \gammadir +jet for 11 $<$ \ETtrig\ $<$ 15 GeV and 15 $<$ \ETtrig\ $<$ 20 GeV, respectively. Right: Jet acoplanarity measurement with \rr\ = 0.5 and 11 < \ETtrig\ < 15 GeV in 0-15\% central \AuAu\ collisions at \sNN\ = 200 GeV. 
    } 
    \label{Fig:GammaJetAuAu}
\end{figure*}

The STAR experiment reports the recoil jet yield ratios between \rr\ = 0.2 and 0.5 (\Rbrtwofive) from direct photon (\gammadir) and \pizero\ triggers. This study is conducted with different trigger transverse energies (\ETtrig), and the results are illustrated in the left side plot of Fig.~\ref{Fig:GammaJetAuAu}. It is evident that the values of \Rbrtwofive\ in 0-15\% central \AuAu\ collisions, at \sNN\ = 200 GeV, are lower than those observed in \pp\ collisions for both \gammadir\ and \pizero\ triggers. This observation serves as a clear and distinct signature of intra-jet broadening in heavy-ion collisions, and contributes to our understanding of jet quenching and the underlying dynamics within  QGP. \\

Jet acoplanarity stands out as another manifestation of jet-medium interaction in heavy-ion collisions. Different physics mechanisms~\cite{DEramo:2018eoy,Mueller:2016gko} can lead to the observations of excess recoil jet yields at $\pi/2$ radian in heavy-ion collisions relative to its \pp\ baseline. The STAR experiment presents such observation in semi-inclusive \gammadir+jet and \pizero+jet measurements with \rr\ = 0.5 and within 11 < \ETtrig\ < 15 GeV. The PYTHIA-8 is used as a baseline for \pp\ data. This is the first observation of jet acoplanarity due to jet-medium interaction in heavy-ion collisions at RHIC. The detailed discussion can be found in Ref.~\cite{YangHP2023}.   

%_______________________ Gamma+jet Fig.5 ______

%%%________ 
{\it Baryon-to-Meson Ratios in Jets}: 
The enhancement of the baryon-to-meson yield ratio in heavy-ion collisions compared to \pp\ collisions is explained by the recombination and coalescence mechanisms of particle production in QGP~\cite{STAR:2006uve}. Whether these mechanisms contribute to the hadronization process in a jet or not has become a focal point in heavy-ion collision experiment. The STAR experiment has conducted the initial measurement of the proton-to-pion yield ratio within a jet, with \rr\ = 0.3, in both \pp\ and \AuAu\ collisions at \sNN\ = 200 GeV. The jet selection involves a hard-core criterion with constituent particles having \pT\ > 3 \gev\ and reconstructed charged jet \pTjet\ > 10 \gev.

\begin{figure*}[htb!]
    \centering
    \includegraphics[width=0.4\textwidth]{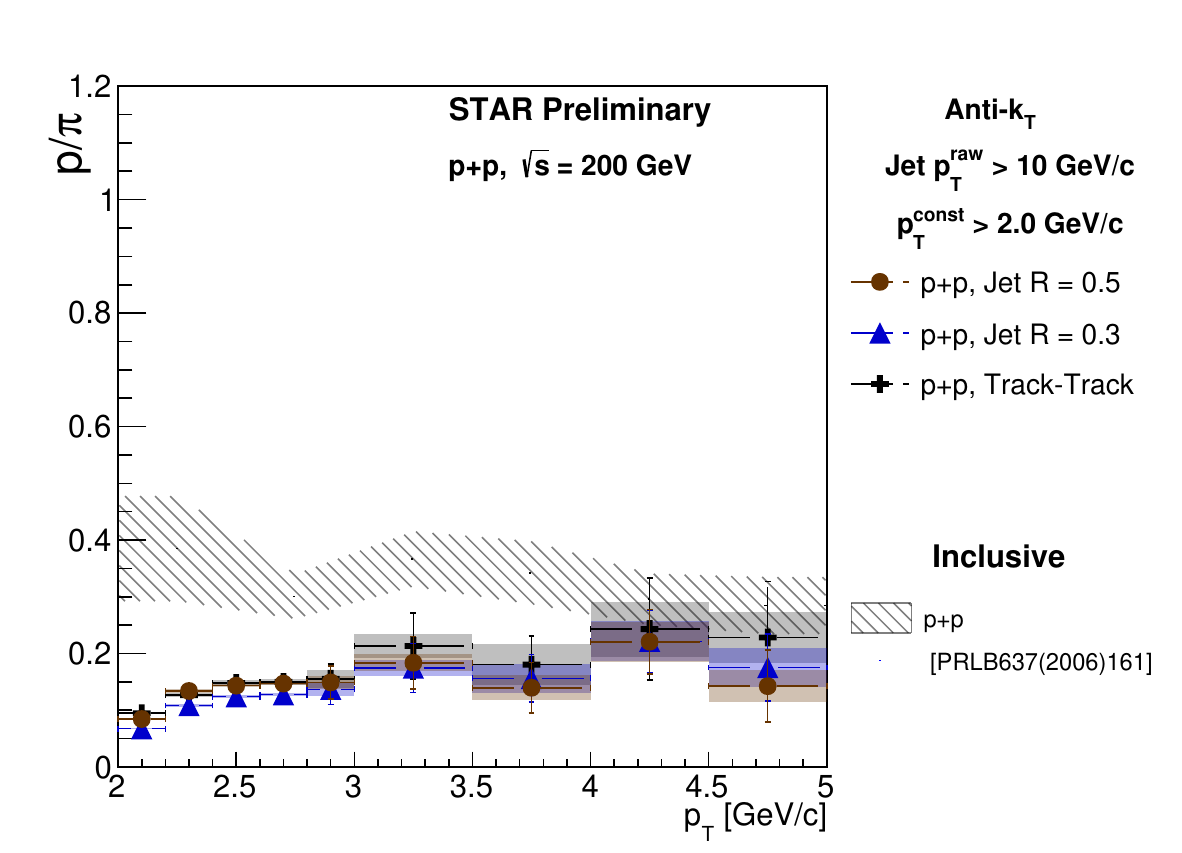}
     \includegraphics[width=0.4\textwidth]{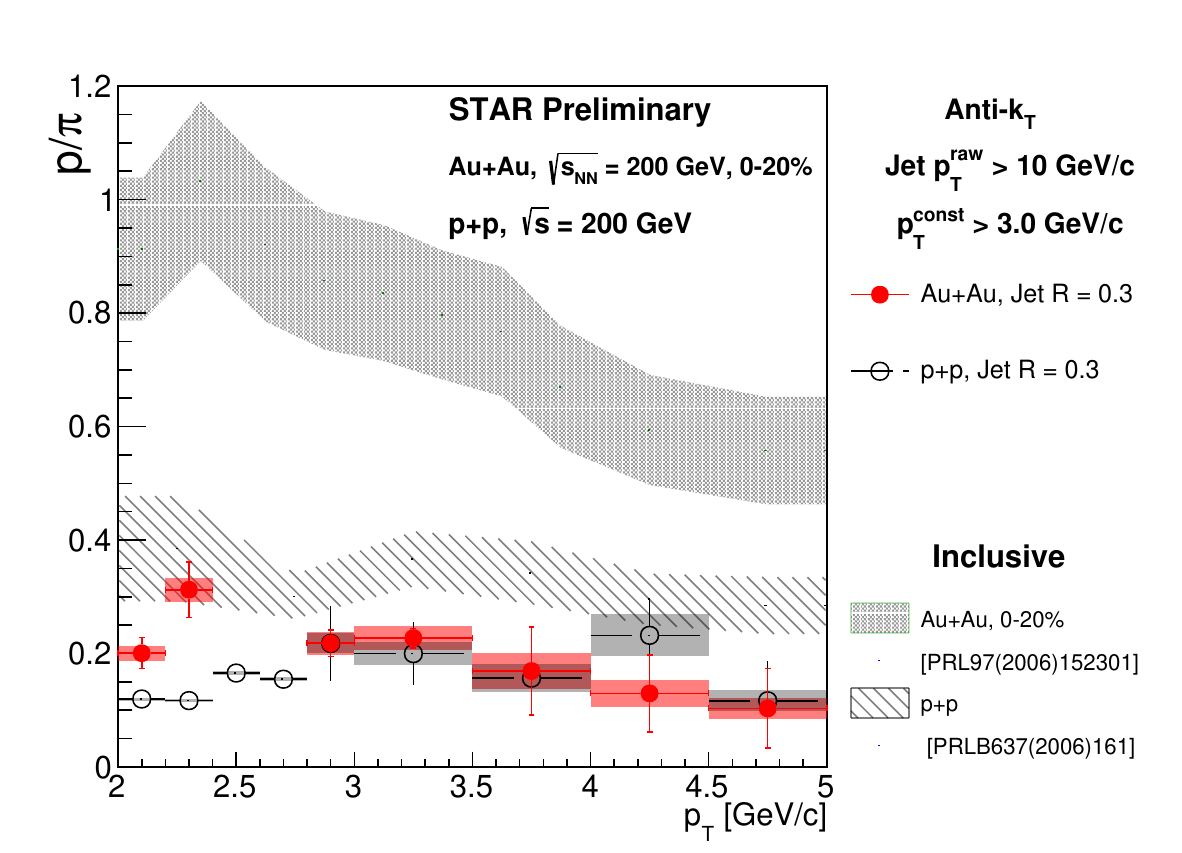}
    \caption{Left: The $(p+\bar{p})/(\pi^{+}+\pi^{-})$  ratio in \pp\ collisions within jets with \rr\ = 0.3 and 0.5. Right: comparison between \pp\ and \AuAu\ for jet \rr\ =0.3 and reconstructed jet \pTjet\ > 10 \gev\ in \pp\ and 0-20\% central \AuAu\ collisions at \sNN\ = 200 GeV.
    } 
    \label{Fig:JetBaryon2Meson}
\end{figure*}

In \pp\ collisions, the $(p+\bar{p})/(\pi^{+}+\pi^{-})$ ratio shows a preference for pion over proton production within jets for different \rr\ shown on the left of Fig.~\ref{Fig:JetBaryon2Meson}. In \AuAu\ and \pp\ collisions at \sNN\ = 200 GeV, for jets with a hard-core selection and radius of 0.3,
the $(p+\bar{p})/(\pi^{+}+\pi^{-})$ ratios are consistent as shown in Fig.~\ref{Fig:JetBaryon2Meson}. Detailed discussion can be found in Ref.~\cite{GabeHP2023}. 
%Similar p/π ratio in Au+Au and p+p with hard core jet selection with constituent pT > 3.0 GeV/c  

%_______________________ Gamma+jet Fig.5 ______

%%%%%%%%%%
{\it Nuclear modification factor and flow in isobar collisions}:
In order to explore the path-length dependence of jet quenching in relatively smaller collision system compared with \AuAu\ collisions, inclusive charged hadron nuclear modification factor (\RAA) and the second-order Fourier coefficient (\vTwo) of charged jets are measured in isobar (\RuRu\ and \ZrZr) collisions at \sNN\ = 200 GeV. 

\begin{figure*}[htb!]
    \centering
    \includegraphics[width=0.37\textwidth]{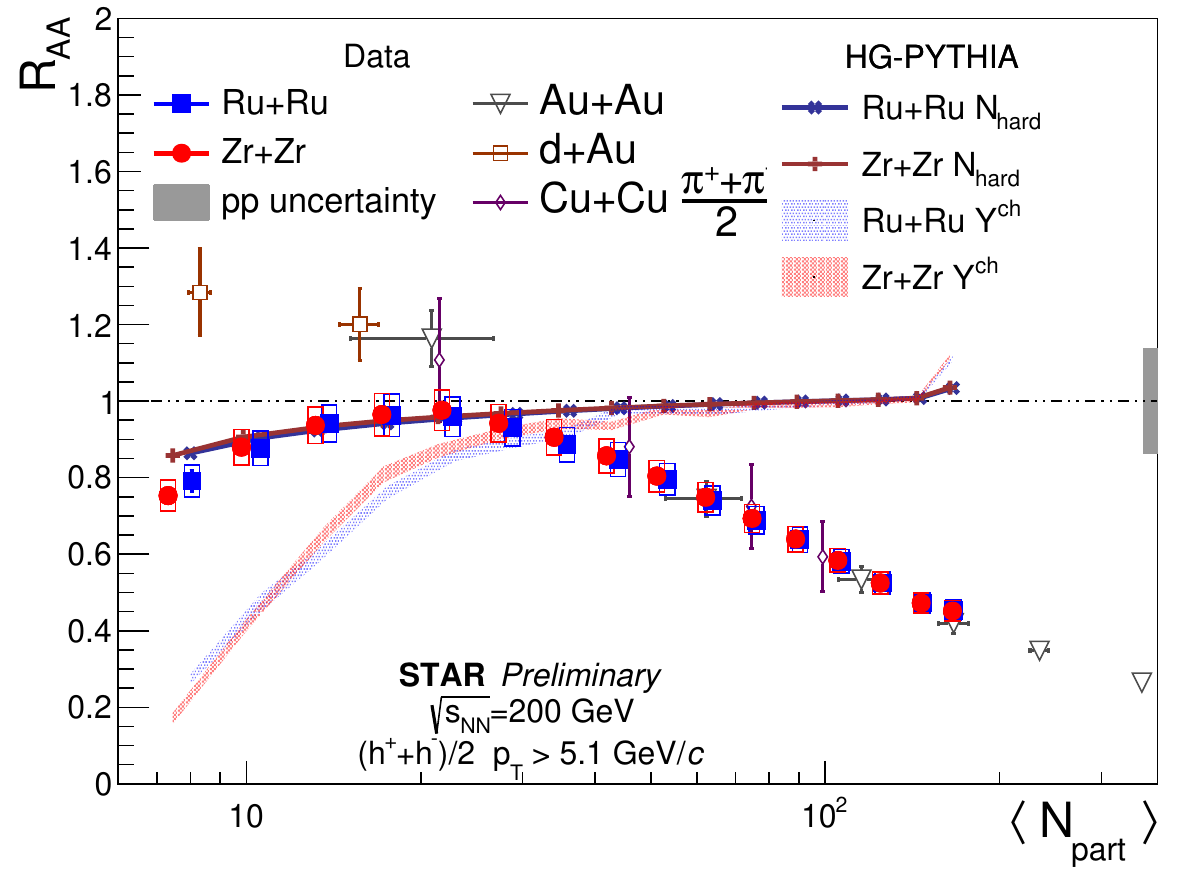}
     \includegraphics[width=0.45\textwidth]{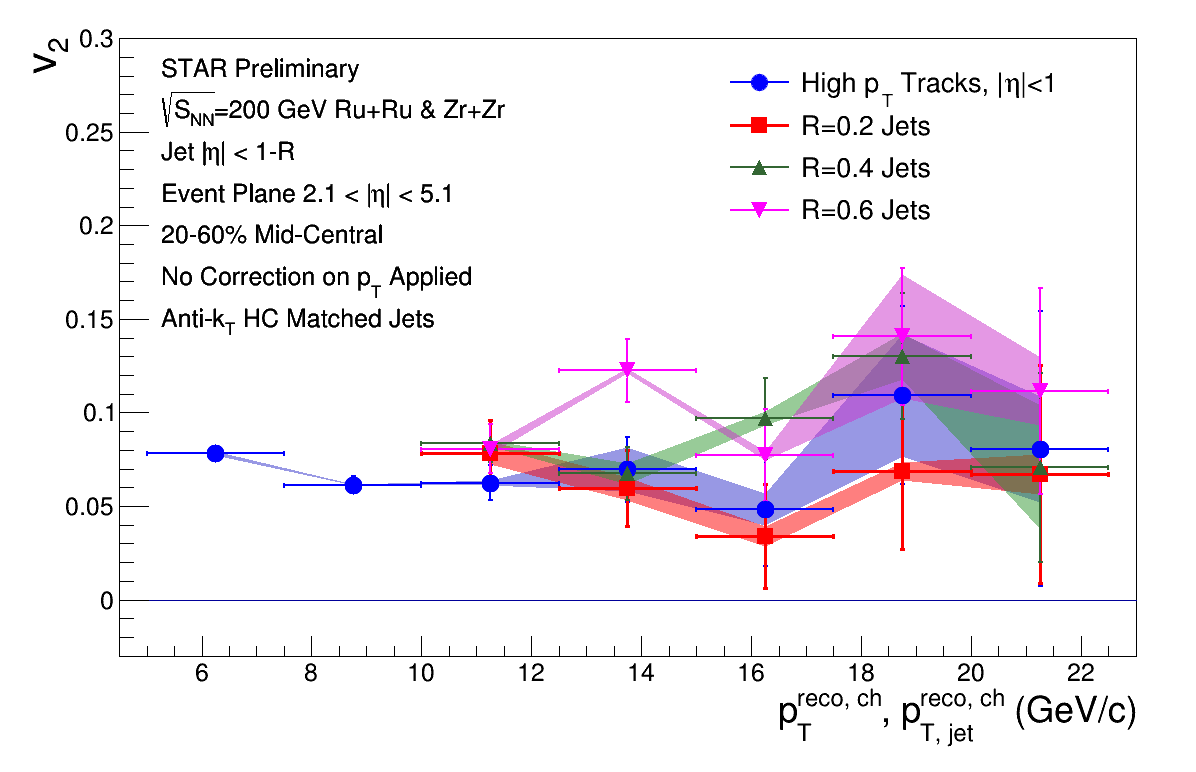}
    \caption{Left: \RAA\ as a function of $\langle  N_{\rm part} \rangle$ for \RuRu\ and \ZrZr\ collisions at \sNN\ = 200 GeV and their comparison with other collisions systems.  Right: Jet $v_{2}$ as a function of jet \pTjet\ in \RuRu\ and \ZrZr\ collisions for different \rr. Hard-core jet selection is applied and no correction is applied along \pT-axis.  
    } 
    \label{Fig:IsobarRAAJetFlow}
\end{figure*}

Figure~\ref{Fig:IsobarRAAJetFlow} shows the inclusive charged hadron \RAA\ measured with \pT\ > 5.1 \gev\ in isobar collisions and compared with other collision systems. A similar suppression is observed at comparable average number of participating nucleons (\Npart), implying energy density in heavy-ion collisions plays the key role to the \RAA\ relative to the initial geometry of the collisions. 

The jet \vTwo\ as a function of reconstructed jet \pT\ in isobar collisions at \sNN\ = 200 GeV is shown in Fig.~\ref{Fig:IsobarRAAJetFlow} for \rr\ = 0.2, 0.4, and 0.5. High-\pT\ jets show non-zero \vTwo\ and no jet \rr\ dependence is observed. In the overlapping kinematic region, the measured \vTwo\ is consistent with that from \sNN\ = 2.76 TeV in Pb+Pb collisions. Detailed discussion can be found in Ref.~\cite{TristanHP2023}.

A measurement of semi-inclusive recoil jets coincidence with a charged hadron within trigger \pT\ between 7 to 25 \gev\ in isobar collisions is also presented~\cite{YangHP2023}. A detailed study is ongoing in this direction to explore jet quenching in smaller collision system. 

%_______________________ Gamma+jet Fig.5 ______

%%%%%%%%%%%%%%%%%%
\section{Heavy-flavor measurements} 

In STAR, the \Jpsi\ measurements in different collision systems and suppression of different $\Upsilon$ states in \AuAu\ collisions are discussed in this section.

%%%%%%%%
{\it \Jpsi\ suppression and flow in isobar collisions}: 
The \Jpsi\ suppression depends on the medium temperature, and is due to the color screening in the QGP. The \Jpsi\ production in heavy-ion collisions is an interplay between the dissociation and regeneration effect; Thanks to the large statistics isobar collisions data, we can test the collision system size dependence of the J/psi suppression at RHIC. 

%In addition, the regeneration and dissociation of \Jpsi\ play important roles to the production of charmonia in heavy-ion collisions, and also depend on the size of collision system. 

\begin{figure*}[htb!]
    \centering
    \includegraphics[width=0.4\textwidth]{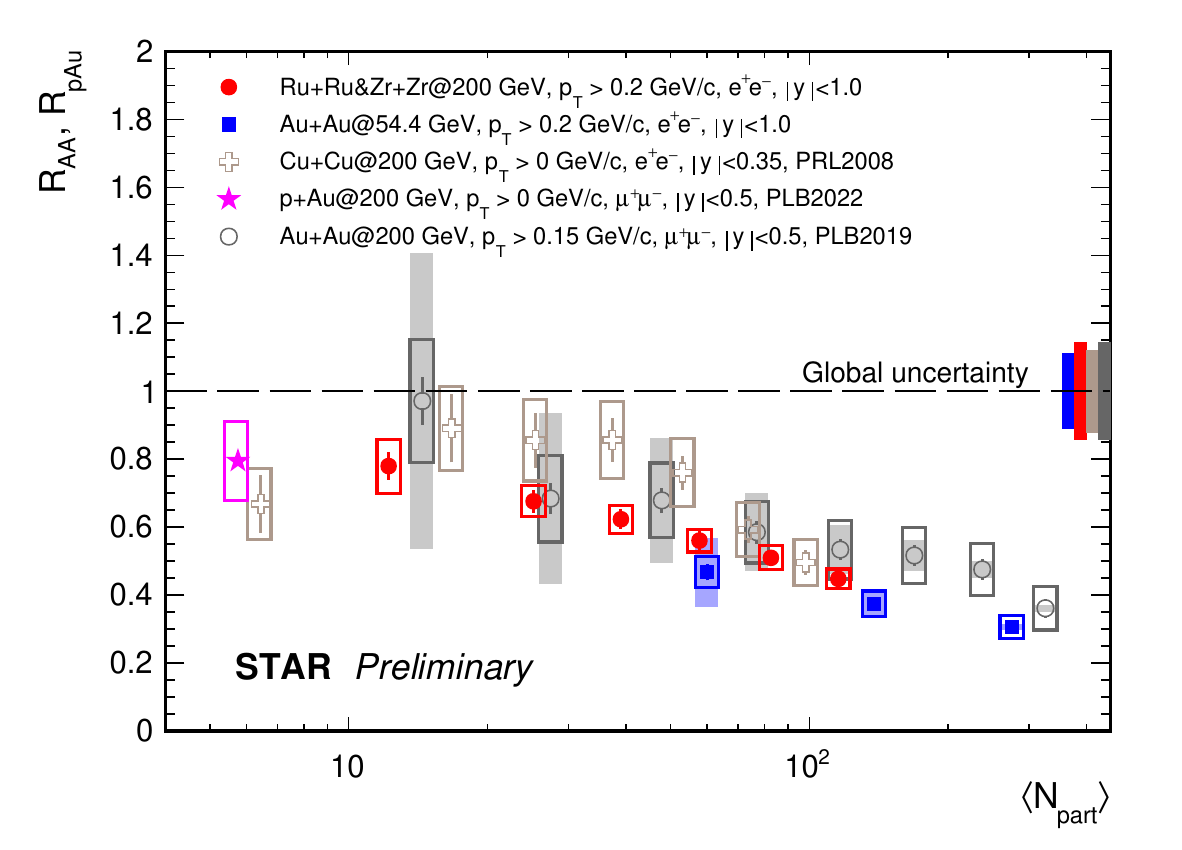}
   \includegraphics[width=0.4\textwidth]{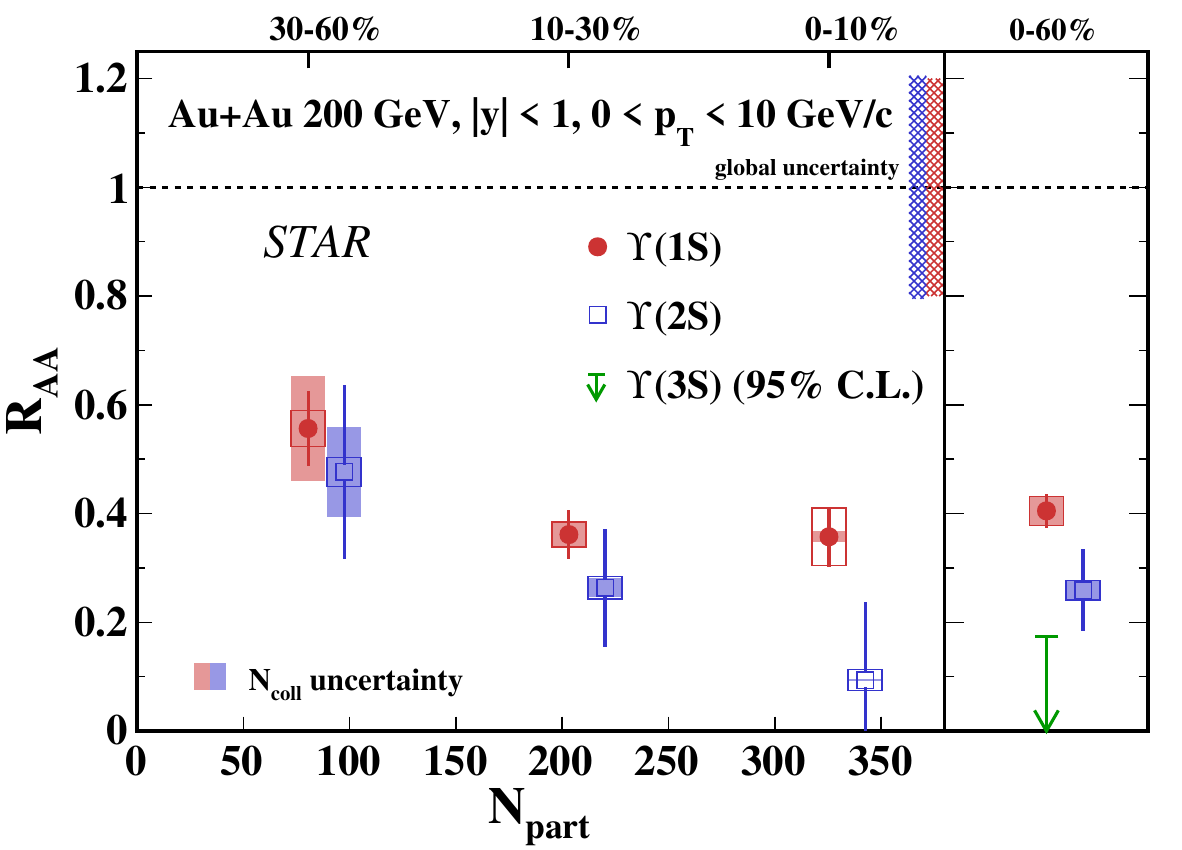}
    \caption{Left: \Jpsi\ \RAA\ as a function of $\langle  N_{\rm part} \rangle$ for \RuRu\ and \ZrZr\ collisions at \sNN\ = 200 GeV, and their comparison with other collisions systems and colliding energies. Right: $\Upsilon$ \RAA\ as a function of \Npart\ in \AuAu\ collisions at \sNN\ = 200 GeV. 
    } 
    \label{Fig:JpsiUpsilonRAA}
\end{figure*}

The \Jpsi\ \RAA\ shows no significant collision system and colliding energy dependence for similar \Npart\ as in the left plot of Fig.~\ref{Fig:JpsiUpsilonRAA}.
%However, the \Jpsi\ \RAA\ is shown in the left plot of Fig~\ref{Fig:JpsiUpsilonRAA} for different collision systems. No significant collision system and energy dependence are observed for \Jpsi\ \RAA\ at similar \Npart. A strong \Jpsi\ suppression is seen at large \Npart. %in \AuAu\ collisions. 
In addition, \Jpsi\ \vTwo\ is also reported in isobar collisions and its comparison with \AuAu\ collisions (plot is not included in these proceedings); No significant \Jpsi\ \vTwo\ is observed at the current level of precision. Detailed discussion can be found in Ref.~\cite{YanHP2023}. 

%%%%%%%
{\it Sequential $\Upsilon$ suppression in \AuAu\ collisions}:
Sequential suppression of bottomonia--bound states of $b$ and $\bar{b}$--in QGP tells about the medium temperature. Different bottomonia states, like $\Upsilon$(1S), $\Upsilon$(2S), and $\Upsilon$(3S), have different binding energies and sizes. The STAR experiment recently published~\cite{STAR:2022rpk} the measurements of $\Upsilon$ production in \AuAu\ collisions at \sNN\ = 200 GeV using both the dielectron and dimuon decay channels. It is observed that $\Upsilon$(2S) and $\Upsilon$(3S) states are significantly more suppressed than $\Upsilon$(1S) state in \AuAu\ collisions relative to \pp\ as shown in the right plot of Fig.\ref{Fig:JpsiUpsilonRAA}. This implies that the QGP temperature, at RHIC, is sufficient to melt excited bottomonia states~\cite{STAR:2022rpk}. 
 %_______________________ Gamma+jet Fig.8 ______

%%%______________
%\section{From QCD in vacuum to hot-dense QCD medium: Big Picture}

%%%%%%%________
\section{Ongoing hard probes physics program at STAR}
The STAR experiment recently underwent an upgrade of its forward detector subsystems, introducing the FTS and FCS components as discussed in Sec.\ref{Sec:STARdetector} and as shown in Fig.~\ref{Fig:STARdetector}. This upgrade not only enhances the capabilities of the Cold QCD physics program but also offers unique opportunities for measuring jets and photons at forward rapidity, thereby benefiting the hard probes physics program at STAR.

\begin{figure*}[htb!]
    \centering          \includegraphics[width=0.6\textwidth]{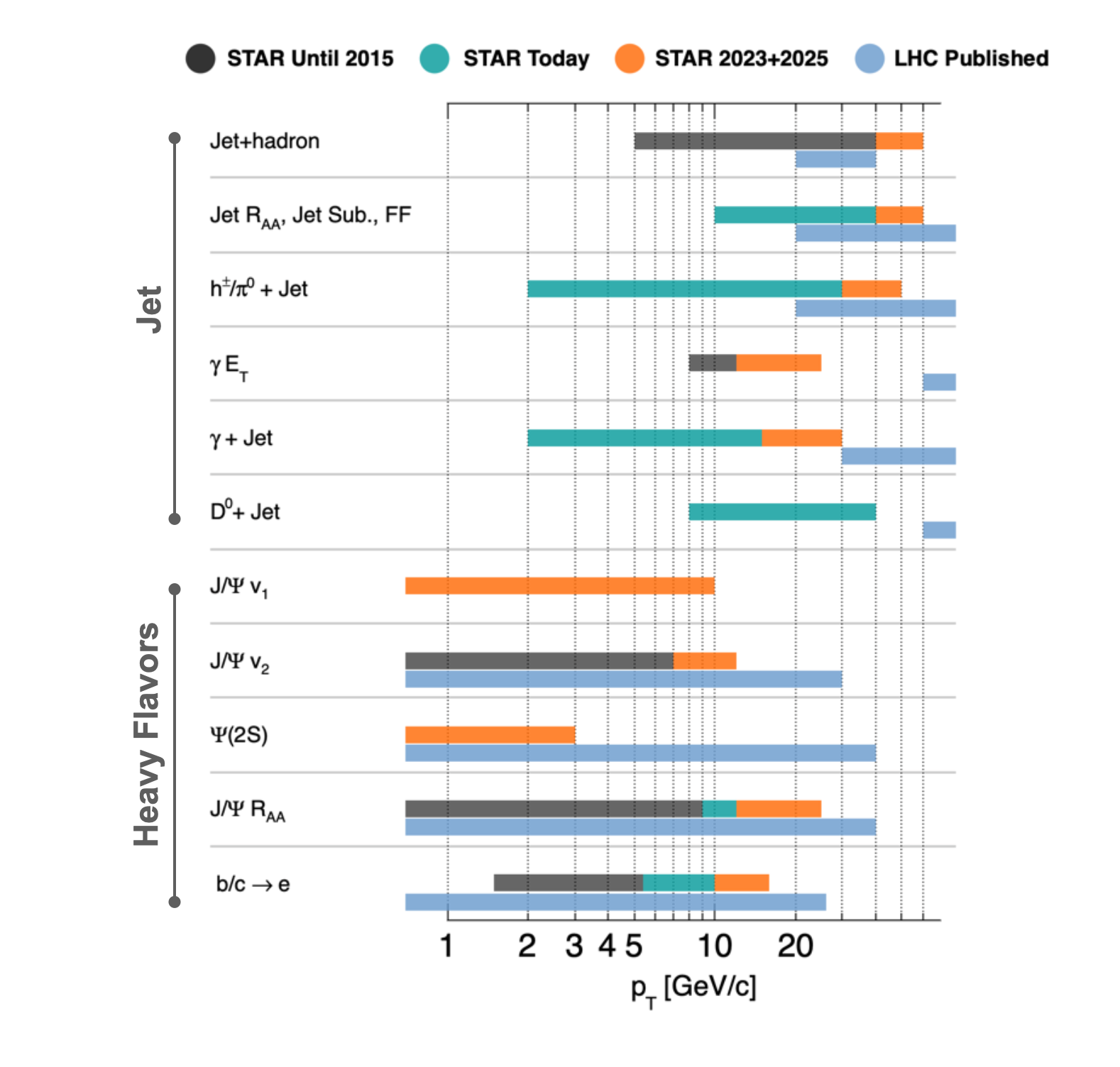}
 \vspace{-6mm}
    \caption{The kinematic coverage of hard probes for precision measurements using both \AuAu\ and \pp\ collisions data  between 2023 to 2025 (Orange band); The black, green, and blue bands represent the STAR measurements untill 2015, current STAR, and LHC published, respectively.}
    \label{Fig:HPkinematic}
\end{figure*}

Furthermore, the STAR experiment has plans to accumulate high luminosity data for precision measurements of various hard probes~\cite{STARBUR}. %Specifically, it is aimed to collect 40 $\rm nb^{-1}$ of \AuAu\ collision data at \sNN=200 GeV in 2023 and 2025. 
Specifically, it is aimed to collect 20 billion minimum bias and 40 $\rm nb^{-1}$ triggered \AuAu\ collisions data at \sNN\ = 200 GeV in 2023 and 2025.
Additionally, data from 235 $\rm pb^{-1}$ of \pp\ collisions and 1.3 $\rm pb^{-1}$ of $p$+Au collisions will be collected in 2024. The kinematic coverage of various hard probes measurements are sketched in Fig.~\ref{Fig:HPkinematic} including the published LHC and RHIC results. This extensive dataset will enable in-depth studies of the micro-structure of the Quark-Gluon Plasma (QGP) through jet and heavy-flavor measurements.

%_______________________ STAR HP plan Fig.9 ______
%\vspace{-10mm}
%%%______________
\section{Summary and outlook}
The STAR experiment covers a broad range of measurements, investigating hard probes to study both QCD in vacuum and the QCD medium at finite temperature. In the context of jet substructure measurements in \pp\ collisions, the interplay between pQCD and npQCD regimes is explored through the vacuum parton shower and a universal scale for parton-hadron transition.

In \AuAu\ collision systems, we report different manifestations of jet-medium interaction, including jet suppression, intra-jet broadening, and recently observed jet-acoplanarity, along with the suppression of bottomonia states. These observations help to understand the microstructure and dynamics of QGP in heavy-ion collisions at RHIC.

For smaller collision systems such as isobar collisions, key observations include comparable levels of inclusive charged hadron and \Jpsi\ suppression as observed in \AuAu\ collisions along with finite jet \vTwo, but not significant \Jpsi\ \vTwo. %This raises questions about the sensitivity of parton energy loss and flow of jets and heavy-flavors to the initial energy density or collision geometry in heavy-ion collisions.
To understand these observations, a detailed study is ongoing to explore the effect of the initial energy density or collision geometry in heavy-ion collisions.

The STAR experiment plans to further explore the properties of QGP through precision hard probes measurements. Ongoing data collection is planned until 2025.


\begin{thebibliography}{99}
%\bibitem{...}

%\cite{Wang:1994fx}
\bibitem{Wang:1994fx}
X.~N.~Wang, M.~Gyulassy and M.~Plumer,
%``The LPM effect in QCD and radiative energy loss in a quark gluon plasma,''
Phys. Rev. D \textbf{51}, 3436-3446 (1995) doi:10.1103/PhysRevD.51.3436[arXiv:hep-ph/9408344 [hep-ph]].
%380 citations counted in INSPIRE as of 26 Jul 2023

%\cite{STAR:2022rpk}
\bibitem{STAR:2022rpk}
B.~Aboona \textit{et al.} [STAR],
%``Observation of sequential $\Upsilon$ suppression in Au+Au collisions at $\sqrt{s_{_\mathrm{NN}}}$ = 200 GeV with the STAR experiment,''
Phys. Rev. Lett. \textbf{130}, no.11, 112301 (2023)
doi:10.1103/PhysRevLett.130.112301
[arXiv:2207.06568 [nucl-ex]].
%11 citations counted in INSPIRE as of 26 Jul 2023

%\cite{Dong:2019byy}
\bibitem{Dong:2019byy}
X.~Dong, Y.~J.~Lee and R.~Rapp,
%``Open Heavy-Flavor Production in Heavy-Ion Collisions,''
Ann. Rev. Nucl. Part. Sci. \textbf{69}, 417-445 (2019)
doi:10.1146/annurev-nucl-101918-023806
[arXiv:1903.07709 [nucl-ex]].

%\cite{Chen:2020vvp}
\bibitem{Chen:2020vvp}
H.~Chen, I.~Moult, X.~Zhang and H.~X.~Zhu,
%``Rethinking jets with energy correlators: Tracks, resummation, and analytic continuation,''
Phys. Rev. D \textbf{102}, no.5, 054012 (2020)
doi:10.1103/PhysRevD.102.054012
[arXiv:2004.11381 [hep-ph]].

%\cite{Komiske:2022enw}
\bibitem{Komiske:2022enw}
P.~T.~Komiske, I.~Moult, J.~Thaler and H.~X.~Zhu,
%``Analyzing N-Point Energy Correlators inside Jets with CMS Open Data,''
Phys. Rev. Lett. \textbf{130}, no.5, 051901 (2023)
doi:10.1103/PhysRevLett.130.051901
[arXiv:2201.07800 [hep-ph]].
%33 citations counted in INSPIRE as of 18 Jul 2023


%\cite{AndrewHP2023}
\bibitem{AndrewHP2023}
Andrew Tamis (for the STAR Collaboration), in these proceedings.

%\cite{Larkoski:2014wba}
\bibitem{Larkoski:2014wba}
A.~J.~Larkoski, S.~Marzani, G.~Soyez and J.~Thaler,
%``Soft Drop,''
JHEP \textbf{05}, 146 (2014)
doi:10.1007/JHEP05(2014)146
[arXiv:1402.2657 [hep-ph]].


%\cite{Chien:2019osu}
\bibitem{Chien:2019osu}
Y.~T.~Chien and I.~W.~Stewart,
%``Collinear Drop,''
JHEP \textbf{06}, 064 (2020)
doi:10.1007/JHEP06(2020)064
[arXiv:1907.11107 [hep-ph]].

%\cite{Dreyer:2018nbf}
\bibitem{Dreyer:2018nbf}
F.~A.~Dreyer, G.~P.~Salam and G.~Soyez,
%``The Lund Jet Plane,''
JHEP \textbf{12}, 064 (2018)
doi:10.1007/JHEP12(2018)064
[arXiv:1807.04758 [hep-ph]].
%140 citations counted in INSPIRE as of 03 Aug 2023

%\cite{MonikaHP2023}
\bibitem{MonikaHP2023}
Monika Robotková (for the STAR Collaboration), in these proceedings.

%\cite{TanmayHP2023}
\bibitem{TanmayHP2023}
Tanmay Pani (for the STAR Collaboration), in these proceedings.

%\cite{STAR:2019yqm}
\bibitem{STAR:2019yqm}
J.~Adam \textit{et al.} [STAR],
%``Longitudinal double-spin asymmetry for inclusive jet and dijet production in pp collisions at $\sqrt{s} = 510$  GeV,''
Phys. Rev. D \textbf{100}, no.5, 052005 (2019)
doi:10.1103/PhysRevD.100.052005
[arXiv:1906.02740 [hep-ex]].

%\cite{Aguilar:2021sfa}
\bibitem{Aguilar:2021sfa}
M.~R.~Aguilar, Z.~Chang, R.~K.~Elayavalli, R.~Fatemi, Y.~He, Y.~Ji, D.~Kalinkin, M.~Kelsey, I.~Mooney and V.~Verkest,
%``pythia8 underlying event tune for RHIC energies,''
Phys. Rev. D \textbf{105}, no.1, 016011 (2022)
doi:10.1103/PhysRevD.105.016011
[arXiv:2110.09447 [hep-ph]].

%\cite{DEramo:2018eoy}
\bibitem{DEramo:2018eoy}
F.~D'Eramo, K.~Rajagopal and Y.~Yin,
%``Moli\`ere scattering in quark-gluon plasma: finding point-like scatterers in a liquid,''
JHEP \textbf{01}, 172 (2019)
doi:10.1007/JHEP01(2019)172
[arXiv:1808.03250 [hep-ph]].
%42 citations counted in INSPIRE as of 26 Jul 2023

%\cite{Mueller:2016gko}
\bibitem{Mueller:2016gko}
A.~H.~Mueller, B.~Wu, B.~W.~Xiao and F.~Yuan,
%``Probing Transverse Momentum Broadening in Heavy Ion Collisions,''
Phys. Lett. B \textbf{763}, 208-212 (2016)
doi:10.1016/j.physletb.2016.10.037
[arXiv:1604.04250 [hep-ph]].
%61 citations counted in INSPIRE as of 26 Jul 2023

%\cite{YangHP2023}
\bibitem{YangHP2023}
Yang He (for the STAR Collaboration), in these proceedings.

%\cite{STAR:2006uve}
\bibitem{STAR:2006uve}
B.~I.~Abelev \textit{et al.} [STAR],
%``Identified baryon and meson distributions at large transverse momenta from Au+Au collisions at s(NN)**(1/2) = 200-GeV,''
Phys. Rev. Lett. \textbf{97}, 152301 (2006)
doi:10.1103/PhysRevLett.97.152301
[arXiv:nucl-ex/0606003 [nucl-ex]].

%\cite{YangHP2023}
\bibitem{GabeHP2023}
Gabriel Dale-Gau (for the STAR Collaboration), in these proceedings.

%\cite{TristanHP2023}
\bibitem{TristanHP2023}
Tristan Protzman (for the STAR Collaboration), in these proceedings.

%\cite{TristanHP2023}
\bibitem{YanHP2023}
Yan Wang (for the STAR Collaboration), in these proceedings.

%\cite{STARBUR}
\bibitem{STARBUR}
STAR BUR at the BNL NPP 2022 PAC Meeting;
https://indico.bnl.gov/event/15148%/attachments/40846/68609/STAR\_BUR\_Runs23\_25\_\_\_2022\%20\%281\%29.pdf

\end{thebibliography}
\end{document}